\documentclass[twocolumn,prd,10pts,nofootinbib]{revtex4-1}
\usepackage[utf8]{inputenc}

\def\mysections#1{{\bf #1.} } 
\usepackage{hyperref}
\usepackage{braket}  
\usepackage[dvips]{graphicx}
\usepackage{hhline}
\usepackage{epsfig,amsmath,amssymb,verbatim,mathrsfs,array,layout,textcomp,amssymb,latexsym,slashed,graphicx,booktabs,color,mathtools}

\newcommand{\beq}{\begin{equation}}
\newcommand{\eeq}{\end{equation}}

\def\beqa{\begin{eqnarray}}
\def\eeqa{\end{eqnarray}}
\def\bea{\begin{eqnarray}}
\def\eea{\end{eqnarray}}

\newcommand{\bv}{\left(\begin{array}{c}}
\newcommand{\ev}{\end{array}\right)}
\newcommand{\bmtwo}{\left(\begin{array}{cc}}
\newcommand{\bmthree}{\left(\begin{array}{ccc}}
\newcommand{\emn}{\end{array}\right)}
\newcommand{\bmtwoc}{\left\{\begin{array}{cc}}
\newcommand{\bmthreec}{\left\{\begin{array}{ccc}}
\newcommand{\emnc}{\end{array}\right\}}
\newcommand{\ba}{\begin{array}}
\newcommand{\ea}{\end{array}}

\newcommand{\BT}{\rule{0pt}{3ex} \rule[-2ex]{0pt}{0pt}}

\def\lsim{\mathrel{\rlap{\lower4pt\hbox{\hskip1pt$\sim$}}
     \raise1pt\hbox{$<$}}}         
\def\gsim{\mathrel{\rlap{\lower4pt\hbox{\hskip1pt$\sim$}}
     \raise1pt\hbox{$>$}}}         

\begin{document}

\title{
Self Destructing Atomic DM
}

\author{Michael Geller${}^{1}$}\email{mic.geller@gmail.com }
\author{Ofri Telem${}^{2\,3}$}\email{t10ofrit@gmail.com}

\affiliation{${}^1$School of Physics and Astronomy, Tel-Aviv University, Tel-Aviv 69978, Israel}
\affiliation{${}^2$ Theory Group, Lawrence Berkeley National Laboratory, Berkeley, CA 94720, USA}
\affiliation{${}^3$ Berkeley Center for Theoretical Physics, University of California, Berkeley, CA 94720, USA}

\begin{abstract}
Self-Destructing Dark Matter (SDDM) is a class of dark sector models in which the collision of a dark sector particle with the earth induces its prompt decay into Standard Model particles, generating unique signals at neutrino detectors. The inherent fragility of SDDM makes its survival from the early universe unlikely,  implying a late time production mechanism. We present an efficient late time production mechanism for SDDM based on atomic rearrangement, the mechanism responsible for muon or anti-proton capture in hydrogen. In this model, an atomic rearrangement process occurs in our galaxy, converting dark atoms into highly excited bound states - our SDDM candidates. While the resulting SDDM is only a small fraction of the dark matter flux, its striking self-destruction signals imply a significant discovery reach in the existing data from the Super-Kamiokande experiment.
\end{abstract}
\maketitle

\section{Introduction}

The nature of dark matter (DM), which makes up 75\% of the matter density in the universe, is still an open question. In the past two decades, direct searches for DM have been guided by the Weakly Interacting Massive Particle (WIMP) paradigm, in which DM is a single particle that interacts weakly with Standard Model (SM). Following the stringent bounds on this scenario from direct and indirect detection, much of the experimental and theoretical focus has recently shifted from WIMP candidates to more broad dark sectors, potentially including sub-GeV DM.

A common thread in DM searches is the direct detection of nuclear or electron recoils from their elastic collisions with incoming DM particles \cite{Goodman:1984dc}. Searches for WIMP-nucleon recoils have been conducted in the Xenon1T \cite{Aprile:2018dbl}, LUX \cite{Akerib:2016vxi}, PandaX-II \cite{Cui:2017nnn}, and CRESST-II \cite{Angloher:2015ewa} detectors, to name a few. The current leading bound on spin-independent WIMP-nucleon elastic scattering was set by the Xenon1T detector, with a minimum of $\sigma=4.1\times10^{-47}\,\text{cm}^2$ at $30\,\text{GeV}$. The limit was set after 278.8 days of data collection, with a fiducial detector volume of 2 tons filled with radio-pure liquid xenon.

In comparison, the fiducial volume of the Super-Kamiokande (Super-K) neutrino detector is 50,000 tons of ultra-pure water \cite{Richard:2015aua}, and so it is tempting to try and harness its large volume to search for DM. Unfortunately, neutrino detectors like Super-K have $\mathcal{O}(\text{MeV})$ thresholds, far above the typical DM-nucleon recoil energy of
\begin{eqnarray}
E_{\text{recoil}}~\sim~\frac{1}{2}r\,m_{\text{DM}}\,v^2_{\text{DM,gal}}~\lsim~\text{keV}\,,
\end{eqnarray}
where $v_{\text{DM,gal}}\sim10^{-3}c$ and $r=4m_{\text{DM}}m_{\text{N}}/(m_{\text{DM}}+m_{\text{N}})^2$ is the efficiency factor of the collision, which peaks for $m_{\text{DM}}=m_{\text{N}}$\footnote{For $m_{\text{DM}}\ll m_{\text{N}}\sim 1\,\text{GeV}$, nuclear recoils are an inefficient way to look for DM, and so electron recoils are used instead \cite{Angle:2011th,Agnes:2018ves,Agnese:2017jvy,Crisler:2018gci,Agnese:2018col}. The typical electron recoil energy is $\mathcal{O}(\text{eV})$.}. 

A caveat to the above no-go argument is self destructing dark matter (SDDM), first introduced in \cite{Grossman:2017qzw}. In essence, SDDM is a meta-stable state in the dark sector, whose interaction with the earth or with  particle detectors can induce its decay. As a result, all of the rest mass of the SDDM is converted to a detectable signal, with an energy $c^2/v^2_{\text{DM}} \sim 10^6$ higher than a typical nuclear recoil signal. This allows for the novel use of high-threshold, large neutrino detectors in the search for this DM candidate. The resulting extraordinary signal - four jets or two simultaneous, highly energetic lepton pairs within the detector - would be a smoking gun for DM self-destruction. The signals of SDDM are very different from those of other DM models that can be probed at neutrino detectors \cite{Davoudiasl:2011fj,Huang:2013xfa,Agashe:2014yua,Berger:2014sqa,Kong:2014mia,Alhazmi:2016qcs,
Kim:2016zjx,Kachulis:2017nci,Dror:2019onn,Dror:2019dib}.

The meta-stability of SDDM, together with its inherent tendency to decay upon scattering, make its early universe production challenging. This was addressed in \cite{Grossman:2017qzw} by suggesting that SDDM could be produced via late time processes in the galaxy. In this work we present a concrete mechanism for the late time production of SDDM out of partially atomic DM. Strikingly, even though SDDM composes only an extremely small fraction of the DM, it can already be discovered in existing neutrino detectors such as Super-Kamiokande.  
 
Our late time SDDM production mechanism is based on the process of atomic rearrangement, analogous to cosmic ray muon capture in hydrogen. This process, first explored by Fermi and Teller in 1947 (\cite{Fermi:1947uv} see also \cite{Wightman:1950zz,Bracci:1979iw,Hydrogen1,Hydrogen2,Hydrogen3,Geller:2018biy}), involves a negative muon and a hydrogen atom rearranging into muonic hydrogen and a free electron. Remarkably,
the typical cross section for this process is the geometric size of the hydrogen atom - much larger than the size of the $(\mu p)$ system. This is an indication that the electron serves as a catalyst for the process. Due to the large impact parameter of the muon and the proton, the resulting muonic hydrogen is formed at an extremely excited state, both in energy and in angular momentum.

We utilize this mechanism in our model by considering a dark sector with heavy $X$ and light $\ell$ fermions, charged under a dark $U(1)$. In the galaxy, $(X\ell)$ atoms are formed, and together with $\bar{X}$ are rearranged into highly excited $(X\bar{X})$ states - our SDDM. These excited $(X\bar{X})$ states are protected from self-annihilation by their high angular momentum, and from spontaneous emission by a small dark photon mass. 

When incident upon the earth, the $(X\bar{X})$ exhibit a typical SDDM phenomenology: they can collide with the nucleus through the dark photon portal and undergo a radiative transition to an unstable, low angular momentum state. Subsequently, they self-annihilate into two or more dark photons, leading to potential striking signals such as two simultaneous lepton or jet pairs in the detector. 

The outline of the paper is as follows. In section~\ref{sec:atomic_rear} we review the physics of atomic rearrangement. In section~\ref{sec:SDDM_general} we present the main ingredients of a dark sector that incorporates the galactic production of SDDM. Next, in section~\ref{sec:toy} we provide a particular implementation of our framework, exploring in section~\ref{sec:dynamics} its early universe and late time dynamics in detail, including the stability of SDDM to self-annihilation, spontaneous emission, collisional de-excitation and inverse rearrangement. In section~\ref{sec:detection} we calculate the discovery reach for our model in the Super-Kamiokande detector, and show how its large fiducial volume allows for a significant discovery reach in the $m_X-\epsilon$ plane.

\section{Atomic rearrangement}
\label{sec:atomic_rear}
At the core of our late time production mechanism for SDDM is the process of atomic rearrangement. This simple mechanism plays an elegant role in SM physics, while it remains mostly overlooked in model building beyond the SM  (see \cite{Kang:2006yd,DeLuca:2018mzn,Contino:2018crt,Geller:2018biy,Mitridate:2017oky} for implementations of rearrangement in the context of a confining gauge group). 
Atomic rearrangement was first studied in a pioneering work by Fermi and Teller \cite{Fermi:1947uv}. In a modern context, rearrangement is what happens when a cosmic $\mu^-$, that has already been slowed down, is incident upon a hydrogen atom, leading to the process:
\begin{equation}
H\,+\,\mu^-~\rightarrow~\left(p^+\,\mu^-\right)\,+\,e^-\, ,
\end{equation}
in which the electron is ionized and the muon is captured. This happens when the muon energy is below the binding energy of hydrogen. In \cite{Fermi:1947uv}, Fermi and Teller found that the cross section for this process is at least geometric,
\begin{equation}
\sigma_{\text{rear}}\,\ge \pi r^2_c~~,~~r_c=0.638\,r_{\text{Bohr}}\, .
\end{equation}
To understand the significance of this result, note that the cross section for simple recombination $p^+\,+\,\mu^-\rightarrow \left(p^+\,\mu^-\right)\,+\,\gamma$ is of order $\frac{1}{\alpha^2 m^2_{\mu}}$, which is a factor of $\left(\frac{m_e}{m_{\mu}}\right)^2$ smaller than rearrangement. In this way, the electron crucially serves as a catalyst for enhanced muon capture.

The physics responsible for the large rearrangement cross section becomes clear in the semiclassical picture once we consider the wavefunction of the electron in the potential of classical, adiabatically moving proton and muon. In the limit when the proton and muon are far away, the electron is simply in the hydrogen ground state around the proton. Conversely, in the limit of zero distance between the proton and the muon, their charge is completely screened, and so there cannot be any bound state for the electron. Evidently, at some critical radius, the two-center potential from the proton and the muon can no longer sustain an electron bound state. Explicit calculation \cite{Fermi:1947uv,Wightman:1950zz} shows that this critical radius is $r_c=0.638\,r_{\text{Bohr}}$. 

From this semiclassical intuition we can construct a dynamical picture of the rearrangement process. Any muon incident upon the hydrogen atom with an impact parameter smaller than $r_c$ will ionize it. If the initial kinetic energy of the muon is low enough, the ionized electron will carry enough energy such that the remaining proton and muon bind into $\left(p^+\,\mu^-\right)$, with a cross section which is at least geometric. In practice, for slow enough muons, the capture happens even at much larger impact parameters due to the focusing effect of the induced hydrogen dipole.  A detailed calculation of the rearrangement process was presented in \cite{Bracci:1979iw} as a function of the temperature $T$, with an anti-proton instead of a muon. There, the cross section was found to be
\begin{equation}\label{eq:rea}
\sigma_{\text{rear}}\,=\,{\left(\frac{T}{5\,E_B}\right)}^{-\frac{16}{25}}\,r^2_{\text{Bohr}} \, ,
\end{equation}
where $E_B=\frac{1}{2}\,\alpha^2\,m_e$. The calculation in \cite{Bracci:1979iw} was semi-classical and adiabatic, in the sense that it assumed a zero kinetic energy for the outgoing electron. There has been a considerable theoretical effort in going beyond this approximation using semi-classical methods \cite{PhysRevA.27.167,PhysRevA.65.052714,Kwong_1989,Sakimoto_2001} as well as fully quantum solutions \cite{Sakimoto_2004,Tong,Sakimoto_low} (For a comprehensive review, see \cite{Cohen_2004}). These further studies have shown that the actual cross section tends to be up to $50\%$ larger than the adiabatic calculation, and that the typical angular momenta of the outgoing bound state are larger as well. To be conservative, we will use \cite{Bracci:1979iw} as our reference for the process, keeping in mind that the full analysis only strengthens our results.

In the model for SDDM presented in the next section, a similar rearrangement mechanism happens in the galaxy, with a dark bound state $\left(X\ell\right)$ playing the role of the Hydrogen atom, while a dark fermion $\bar{X}$ plays the role of the muon/anti-proton. As we demonstrate in Section~\ref{sec:xxbarprod}, this process effectively converts any existing $(X \ell)$ into a free $\ell$ and an $\left(X\bar{X}\right)$, which is our SDDM candidate.

\section{ Cosmologically Viable SDDM}
\label{sec:SDDM_general}
\subsection{Minimal Components of a Dark Sector with Self-Destruction}\label{sec:components}

In this paper we seek to realize self-destructing dark matter as a dark positronium-like state, as in \cite{Grossman:2017qzw}. For that we need a symmetric abundance of dark fermions $X,\,\bar{X}$, which can form bound states under a dark $U(1)_D$. In the next sections we present an efficient mechanism for the creation of these $(X\bar{X})$ bound states with particularly high angular momentum ($L\sim30$).

As in \cite{Grossman:2017qzw}, the high-L $(X\bar{X})$ are cosmologically stable by virtue of their high angular momentum, i.e. to self-annihilate they need to penetrate through a large centrifugal barrier. Additionally, we will show that these states are stable against de-excitation due to a mass for the binding dark photon $A_D$. When this mass is significantly larger than the energy difference between adjacent energy levels in the $(X\bar{X})$ states, the $(X\bar{X})$ can only de-excite via highly forbidden transitions, effectively rendering them cosmologically stable.

\begin{figure*}[t]
\begin{center}
\includegraphics[width=10cm]{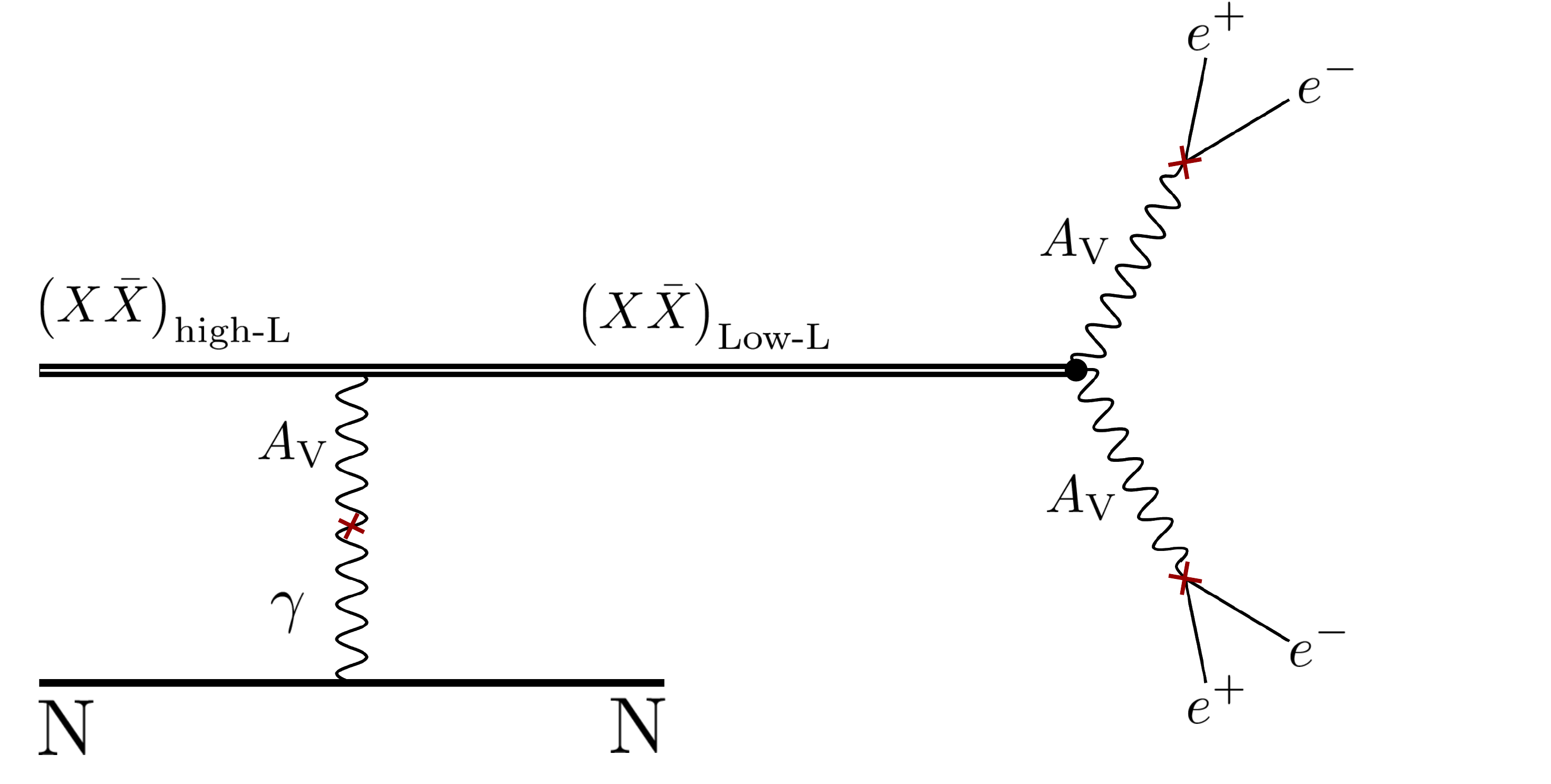}
\caption{\label{fig:SDDMcartoon}
Self-destruction phenomenology. The stable, high-L $(X\bar{X})$ state collides with the SM nucleus through the $A_V$ portal, transitioning to unstable, low-L $(X\bar{X})$. Subsequently, the unstable $(X\bar{X})$self-annihilates into a pair of back-to back $A_V$s, which in turn go to $e^\pm$ pairs (they can also go to jets).
}
\end{center}
\end{figure*}

Next, to allow for the $(X\bar{X})$ to self-destruct (transition to a low-L state and self-annihilate) in the detector, we need a mediator, which we take to be a massive dark photon $A_V$ mixing with the SM photon. The mediator gives rise to the self-destruction process depicted in Fig.~\ref{fig:SDDMcartoon}, namely, a stable, high-L $(X\bar{X})$ state collides with the SM nucleus through the $A_V$ portal, transitioning to an unstable, low-L $(X\bar{X})$ state. Subsequently, the unstable $(X\bar{X})$ decays to two back-to-back dark photons $A_V$, each one flying a macroscopic distance before decaying to SM jets/lepton pairs through its mixing with the SM.

The essential ingredients for SDDM phenomenology are summarized in the top part of table~\ref{tab:SD}. These are the same states that were included in the original SDDM proposal \cite{Grossman:2017qzw} (the motivation for including two $X$ species will be explained in section~\ref{sec:toy}) . 

\subsection{Outline of the SDDM Production Mechanism}\label{sec:overview}

%
%

Our dark sector consists of the essential ingredients of a self-destructing dark sector\footnote{Here we only consider angular momentum protected SDDM. Other possibilities have been explored in \cite{Grossman:2017qzw}, potentially with very different cosmology.} described in the previous section, namely the fermions $X,\,\bar{X}$ and the dark photons $A_D$ and $A_V$. Additionally, we include a light fermion $\ell$, whose role is to catalyze the production of high-L $(X\bar{X})$ in the galaxy, as described in Section~\ref{sec:xxbarprod}. The overall matter content is detailed in table~\ref{tab:SD}.

\begin{table}[h]
\centering
\begin{tabular}{|c|c|c|}
\hline
 \BT           \textbf{Field} $\mathbf{(q_D,q_V)}$~&~\textbf{Description}~~~&~\textbf{Mass} \\\hline
 \BT       $X_{1,2}\,(\pm 1,\mp 1)$  ~&~make up $(X\bar{X})$~&~$1-100~\text{GeV}$~~\\ \hline
 \BT       $A_D$  (0,0)     ~&~binds $(X\bar{X})$~~~&~$1-100~\text{keV}$~~\\ \hline
 \BT       $A_V$ (0,0)  ~&~mediator to the SM~~~&~$0.1-10~\text{GeV}$~~\\ \hline
 \BT       $\ell$  (1,0)     ~&~catalyzes $(X\bar{X})$ production~~&~$1-100~\text{MeV}$~~\\ \hline
 \end{tabular}
\caption{The matter content in a cosmologically viable dark sector with with self-destructing phenomenology. The numbers in parentheses are the $U(1)_D$ charge $q_D$ and the $U(1)_V$ charge $q_V$}\label{tab:SD}
\end{table}
 
In this scenario, the universe is initially populated by the photons $A_V,\,A_D$ and  the fermions $X,\,\bar{X},\,\ell,\,\bar{\ell}$. We postulate a primordial asymmetry between $X$ and $\ell$, whose origin we leave unspecified. 

As we will see in our concrete model, the $X,\,\bar{X}$ freeze-out in the early universe and end up with a mostly  symmetric relic abundance. In contrast, the lighter $\ell$ fermions annihilate much more efficiently, and so below $m_\ell$  the $\bar{\ell}$ population essentially disappears and we are left with an asymmetric abundance of $\ell$ only. 
In total, the dark sector today contains a subcomponent of $X,\,\bar{X}$ and $\ell$ fermions charged under a short range $U(1)_D$ interaction (the range of the $U(1)_V$ is negligible due to the large mass of $A_V$). The abundances are dynamically set such that the universe is neutral under the dark $U(1)_D$. Remarkably, these simple components naturally give rise to $(X\bar{X})$ SDDM in the following manner:

\begin{enumerate}

\item \textit{Galactic production of $(X\bar{X})$:}~In the relatively higher density environment of the galaxy, recombination is turned on, followed by atomic rearrangement. This leads to an ongoing formation of $\left(X\ell\right)$ atoms that are later converted to $\left(X\bar{X}\right)$ by capturing a free $\bar{X}$ and emitting an $\ell$. As we will see in Section~\ref{sec:xxbarprod}, the $\left(X\bar{X}\right)$ is generated in highly excited states with very large angular momentum.

\item \textit{$(X\bar{X})$ stability:}~The $\left(X\bar{X}\right)$ are cosmologically stable by virtue of their high angular momentum and the mass of the dark photon. They are mainly destroyed by the inverse process to atomic rearrangement, and their lifetime is found to be comparable to the age of the universe. The $\left(X\bar{X}\right)$  are constantly produced and destroyed throughout the history of the galaxy, resulting in a small flux of $\left(X\bar{X}\right)$  through the earth.

\item \textit{SDDM detection:}~
When incident upon the detector, the $\left(X\bar{X}\right)$ acts as bona-fide SDDM, i.e. it can collide with nuclei through the $A_V$ portal and subsequently self-annihilate into a pair of $A_V$, which then decay into jets/lepton pairs. In this way, all of the rest mass of the $(X\bar{X})$ is converted into relativistic SM particles.

\end{enumerate}

\section{A Concrete Model}\label{sec:toy}

Here we present a concrete model which implements the cosmological history outlined in the previous section. The matter content of the model is presented in table~\ref{tab:SD}.

As can be seen from the table, the heavy fermions $X_{1,2}$ and the light fermions $\ell$ are charged under $U(1)_D$, whose photon $A_D$ has a $\mathcal{O}(1-100\,\text{keV})$ mass. This prevents long range interactions in the dark sector (Compton wavelength $<\text{nm}$), but allows for $(X_i\ell)$ and $(X_i\bar{X}_i)$ bound states to form. Additionally, the $X_{1,2}$ are charged under a $U(1)_V$ mediated by the $\mathcal{O}(0.1-10\,\text{GeV})$ dark photon $A_V$. This heavy photon mixes with the SM photon, and so acts as the portal for direct detection.

We consider two mass degenerate `flavors' $X_{1,2}$ in order to prevent the mixing of $U(1)_D$ with $U(1)_V$ and $U(1)_{\text{E\&M}}$. This sort of mixing would have been in tension with the constraints on the production of $\ell$s in SN1987a \cite{Chang:2018rso}.
The mixing is forbidden by a discrete global symmetry acting as $X_1 \leftrightarrow X_2,\,A_V\to -A_V$, with the light fermion $\ell$ and the dark photon $A_D$ left invariant. We denote this symmetry $\mathcal{C}_{\leftrightarrow}$, due to the similarity with charge conjugation.  The charges and mass ranges for the different particles are presented in table~\ref{tab:SD}.

The mass of the binding dark photon $A_D$ has to be light enough to allow $(X_i\ell)$ bound states to form, but heavy enough to forbid the spontaneous de-excitation of high-L $(X_i\bar{X}_i)$, 
\begin{equation}\label{eq:dphrange}
f_{{X\bar{X}}}\,\frac{1}{2}\alpha_D^2 \,m_\ell  < m_{D} <   \frac{1}{2}\alpha_D^2\, m_\ell\, .
\end{equation}
The upper bound allows the  recombination of  $(X_i\ell)$ atoms, while the lower bound is from requiring $A_D$ to be heavier than several energy splittings of $(X_i\bar{X}_i)$. The exact factor $f_{X\bar{X}}<1$ is calculated in Eq.~\ref{eq:stability_frac}. 



Throughout most of the text we set
\begin{equation}
r_{X\ell}~\equiv~\frac{m_X}{m_\ell}~=~\frac{m_p}{m_e}~=~1800\, ,
~\alpha_V = \alpha_D = 0.1\,.
\end{equation}
We choose our benchmark mass ratio to be the same as the mass ratio between the SM proton and electron, so that we can easily extrapolate from the SM results for $\bar{p}$ capture in hydrogen \cite{Bracci:1979iw}.
\quad\\
\quad\\
The $U(1)_V$ photon is heavy, $\alpha_V m_X<m_V<m_X$, and so it does not significantly affect the potential within the $(X_i\ell)$ and $(X_i\bar{X}_j)$ bound states. Instead, the $U(1)_V$ photon serves as a mediator for $(X_i\bar{X}_i)$ self-annihilation into the SM. Accordingly, $U(1)_V$ is mixed the SM hypercharge, while $U(1)_D$ is not mixed at tree level\footnote{The kinetic mixing of $A_V$ and $A_{SM}$ breaks $\mathcal{C}_{\leftrightarrow}$, but leaves an unbroken $\mathcal{C}_{\leftrightarrow} \mathcal{C}_{\text{SM}}P$ that forbids the mixing with $A_D$.},
\begin{equation}
L = \frac{1}{4} F^2_D + \frac{1}{4} F^2_V + \epsilon_{V} F_V F_{EM} +  m_V^2 A^2_V + m_D^2 A^2_D\, .  
\end{equation} 
By virtue of our $\mathcal{C}_{\leftrightarrow}$ symmetry, there is no loop level generation of a mixing term $\epsilon_{D} F_D F_{V}$. If this symmetry is broken and the $X_1$ and $X_2$ have a small difference in mass $\delta m$, then this mixing is generated as $\epsilon_D \sim  \frac{2}{9\pi^2} \alpha_D  \alpha_V \frac{\delta m }{m_X}$. As the the light fermions $\ell$ are in the range of masses that can be produced in core-collapse supernovae, this mixing is severely constrained by the SN1987A cooling constraints. Additionally, if $\mathcal{C}_{\leftrightarrow}$ is broken, the $A_V$ dark photons produced in the self-destruction process have a large probability to decay invisibly to an $\ell^+\ell^-$ pair, suppressing our direct detection signal. These effects are negligible as long as $\epsilon_D\ll \epsilon_V$ and  $\epsilon_D  \epsilon_V \lsim 10^{-10}$,  where the discovery reach on $\epsilon_V$ we get is around $10^{-6}-10^{-4}$. Both conditions hold if $\frac{\delta m }{m_X}\lsim 10^{-2}$. 

\section{Early universe and late time dynamics}\label{sec:dynamics}
\subsection{The Origin of $X,\,\bar{X}$ and $\ell$}

We assume that the dark sector is UV-completed such that it is initially in thermal equilibrium with the visible sector after reheating, but decouples from it below the mass $m_V$ of the heavy dark photon $A_V$ (recall that the binding photon $A_D$ does not mix with the SM photon). 

The cosmological abundance of $\ell$ is determined by a primordial asymmetry, analogous to the one responsible for baryogenesis in the SM. This asymmetry takes the form:
\begin{eqnarray}
 2 \,n\left(\bar{X}_1\right) -  2\, n\left(X_1\right)&=&2 \,n\left(\bar{X}_2\right)-2\, n\left(X_2\right) \nonumber\\ &=& n\left(\ell\right) - n\left(\bar{\ell}\right)   = \eta s \,,
\end{eqnarray}
where $s$ is the entropy density and $\eta$ is the asymmetry. We can see that the $U(1)_D$ charges are balanced between $X_{1,2}$ and $\ell$, and so there is no charge excess. In the following sections we drop the $1,2$ labels from $X$, and refer to $X_{1,2}$ collectively as $X$. For a small enough asymmetry $\eta$, the symmetric component of $\Omega_X$ is a simple thermal relic and is given by
\begin{eqnarray}
\label{eq:omega_symm}
\Omega_{\text{symm}} \, h^2~&\sim&~10^{-7} \left(\frac{m_{X}}{\text{GeV}}\right)^2\left(\frac{0.1}{\alpha_D}\right)^2\,.
\end{eqnarray}
This equation holds long as $\Omega_{\text{asymm}}\ll\Omega_{\text{symm}}$, which is the case we consider, and we assume that the two sectors are in thermal equilibrium when $X, \bar{X}$ freeze out.  For the light $\ell$s, the annihilation is much more efficient, and only the asymmetric component of $\ell$ remains.
\\
Once the temperature drops below the recombination temperature, $(X\ell)$ atoms are formed. When the $(X\ell)$ meet a free $\bar{X}$, they can undergo a rearrangement process
\begin{eqnarray}\label{eq:rearr_proc}
 (X\ell)\,+\,\bar{X}~\leftrightarrow~(X\bar{X})\,+\,\ell\, ,
\end{eqnarray}
as we described in detail in the section~\ref{sec:atomic_rear}. The generated $(X\bar{X})$ are highly excited, as we will show, and can potentially de-excite and self-annihilate. We comment on the ensuing early universe dynamics in the next section. Here we simply make sure that it does not lead to a second phase of annihilations \cite{Kang:2006yd,Boddy:2014qxa,Geller:2018biy} which further reduces the symmetric abundance of $X$, $\bar{X}$. This is easily guaranteed by keeping $n_\ell< \frac{H(T_{\text{BSF}})}{\sigma_{\text{rear}}v_{X}}$, such that 
\begin{eqnarray}
\Gamma_{\text{rear}}~=~n_{(X\ell)}\sigma_{\text{rear}}v_{X}~<~n_{\ell}\sigma_{\text{rear}}v_{X}~<~H(T_{\text{BSF}})\,,\nonumber\\
\end{eqnarray}
where $H(T_{\text{BSF}})$ is Hubble at the time of the bound state dynamics. Using the cross section given in Eq.~\ref{eq:rea}, we find the following upper bound on the $\ell$ abundance
\begin{eqnarray}
\label{eq:asym_abundance}
&&\Omega_{\ell}\, h^2 = r_{X\ell}\,\Omega_{\text{asymm}}\, h^2 \lsim  10^{-8} \nonumber\\
&&\left(\frac{m_X}{\text{GeV}} \right)^2  \sqrt{\frac{m_p/m_e}{r_{X\ell}}} \frac{0.1}{\alpha_D}\, .
\end{eqnarray}

Having included light fermions and a light dark photon in our sector, we now turn to the relevant bounds from $\Delta N_{\text{eff}}$. Here we assume that $A_D$ decays out of equilibrium to light unspecified states which were not previously reheated\footnote{To ensure the stability of our SDDM, we assume $A_D$ decays through long-lived intermediate states with mass of the same order as $m_D$. The de-excitation of $(\bar{X}{X})$ through an off-shell $A_D$ is therefore highly suppressed(see section~\ref{sec:stability}). }.  At BBN, the number of relativistic degrees of freedom in the dark sector is $g^{D}_*~=~7.5$, including the light fermions $\ell$ and the binding dark photon $A_D$. To be consistent with the stringent bounds on $\Delta N_{\text{eff}}$ from BBN \cite{Fields:2019pfx} and the CMB \cite{Aghanim:2018eyx}, our dark sector has to be colder than the SM at BBN and CMB times. The contribution from the dark sector to $\Delta N_{\text{eff}}$ is
\begin{equation}
\Delta N_{\text{eff}}~=~\frac{g^{D}_*}{0.46}\,{\left(\frac{T_{\text{SM}}}{T_{\text{dark}}}\right)}^{-4}\,.
\end{equation}

Assuming that the two sectors decouple above the QCD phase transition, i.e. taking $m_V > T_{\text{QCD}}$, 
we get $\Delta N_{\text{eff}} \sim 0.41$. Such a high value can be motivated by the Hubble tension \cite{Riess_2018, Aghanim:2018eyx}, as the combined fit of CMB and the local SHOES measurement \cite{Riess_2018} favors $N_{\text{eff}}=3.27\pm 0.15$. Should the tension disappear without any change in the CMB analysis we would need to satisfy the CMB+BBN $95\%$ C.L. bound on $\Delta N_{\text{eff}}\lsim 0.3$ which would require $T_{\text{SM}}/T_{\text{dark}}>2.71$. This temperature ratio may be obtained by a late time entropy dump into the SM from a decay of a heavy new particle, after the decoupling of the two sectors and before BBN. We choose to remain agnostic about the exact nature of this late time entropy dump, which is unconstrained by current observations, and is overall orthogonal to our work.

\subsection{Early universe bound state dynamics}
Here we briefly review the complex (dark-) recombination dynamics in our model. Dark recombination in our model begins at $T^D \lsim T^D_{\text{rec}}\equiv\frac{1}{2}\,\alpha^2_D \,m_\ell$, when $(X \ell)$ atoms are allowed to form. This triggers a complicated dynamics which involves the following processes:
\begin{itemize}
\item Recombination/ionization:
\begin{eqnarray}
X+\ell~\leftrightarrow~(X\ell)\,+\,A_D
\end{eqnarray}
\item Atomic rearrangement/inverse rearrangement:
\begin{eqnarray}
 (X\ell)\,+\,\bar{X}~\leftrightarrow~(X\bar{X})\,+\,\ell
\end{eqnarray}
\item De-excitation and self-annihilation of $(X\bar{X})$
\begin{eqnarray}
 &&(X\bar{X})_{\text{high-L}}\,+\,(X\,\text{or }\,\bar{X})~\leftrightarrow~(X\bar{X})_{\text{low-L}}\,+\,(X\,\text{or }\,\bar{X})\nonumber\\
 &&(X\bar{X})_{\text{low-L}}~\rightarrow~(A_V\,\text{or } A_D) + (A_V\,\text{or } A_D)\, .
\end{eqnarray}
\end{itemize}
We leave the study of this complicated early universe dynamics for future work, and instead make the most conservative assumption that no $(X\ell)$ or $(X\bar{X})$ states remain from the early universe, and so the only surviving states are free $X,\,\bar{X}$ and $\ell$. The relic abundances of $\{X,\bar{X},\ell\}$ remain unaffected by this discussion as long as Eq.~\ref{eq:asym_abundance} is satisfied.  



\subsection{$(X\bar{X})$ production in the galaxy}\label{sec:xxbarprod}
We've seen above that the early universe abundance of $(X\ell)$ atoms is set dynamically by a complicated freezout process that involves recombination, atomic rearrangement and its inverse, and de-excitation of $(X\bar{X})$. Importantly, due to the low relic abundance of $\ell$, the abundance of free $X,\,\bar{X}$ is not affected by this complicated dynamics. Instead of solving the coupled Boltzmann equations for the $(X\ell)$ freezout process, we simply make the conservative assumption that all the $\ell$s in the galaxy start as free, and only later recombine into $(X\ell)$ atoms. 

Before diving into the galactic recombination dynamics, let us comment on the galactic density profiles for our dark sector components. For simplicity we assume in this paper that the galactic density profiles for $X,\,\bar{X}$ and $\ell$ follow the NFW profile of \cite{Nesti:2013uwa}, scaled by the appropriate ratio of abundances:
\begin{eqnarray}
n^{\text{total}}_{\ell}~&=&~\frac{1}{m_X}\frac{\Omega_{\text{asymm}}}{\Omega_{\text{DM}}}\,\rho^{\text{NFW}}_{\text{\text{gal}}}\nonumber\\
n_{\bar{X}}~=~n_{X}~&=&~\frac{1}{m_X}\frac{\Omega_{\text{symm}}}{\Omega_{\text{DM}}}\,\rho^{\text{NFW}}_{\text{\text{gal}}}\,.
\end{eqnarray}
Where $n^{\text{total}}_{\ell}$ is the total number of $\ell$ particles, bound and free. We have checked all our results with the Burkert distribution from  \cite{Nesti:2013uwa} as well, and found only a weak sensitivity of our results on the assumed profile. Moreover, the NFW profile gives the more conservative results, and so we use it for the calculation of the all the rates.  
We further assume that the profiles of the produced states - $(\bar{X}X)$ and $(X\ell)$ quickly relax to the appropriately scaled profiles. An alternative assumption, that the density of the bound states is governed by local dynamics gives similar results, since the local DM density $\rho_{\text{local}}\simeq 0.3~\text{GeV/cm}^{3}$ is close to the rms value of both the Burkert and NFW distributions in \cite{Nesti:2013uwa}. 
Finally, we checked that the cooling of the species $X,\bar{X},\ell$ is inefficient for gravitational collapse, due to the low abundance of the light species $\ell$ \cite{Fan:2013yva}.  We leave a more detailed analysis of the galactic dynamics in this scenario for future work. 

In contrast with the early universe recombination of $p$ and $e$ in the SM, the galactic recombination of $(X\ell)$ is governed by direct recombination to the ground state. This is because the mean free path for ionization of the outgoing dark photon is larger than the size of the galaxy. Therefore, the production rate for $(X\ell)$ is given by:
\begin{eqnarray}
\label{eq:rec_rate}
&&\frac{d\Gamma_{\text{rec}}}{dV}~=~n_{X}\, n^{\text{free}}_{\ell}\,\sigma_{\text{rec}}\,v_{\ell}~=~8.6\times10^{-26}~\text{cm}^{-3}\text{s}^{-1}\nonumber\\
&&{\left(\frac{ n_X n^{\text{free}}_{\ell}}{\braket{n_X n^{\text{total}}_\ell} }\right)}\,{\left(\frac{r_{X\ell}}{m_p/m_e}\right)}\,{\left(\frac{10^{-3}\,c}{v_{\text{\text{gal}}}}\right)}\, ,
\end{eqnarray}
where $n^{\text{free}}_\ell$ and $n_X$ are the number densities of the free $\ell$ and $X$, and the cross section $\sigma_{\text{rec}}$ for direct recombination was calculated in \cite{berestetskii1982quantum}. We assume that all the $\ell$s are thermalized so that $v_\ell=\sqrt{r_{X\ell}}\,v_{\text{\text{gal}}}$. To verify this assumption, we estimate the mean free time for the $\ell$s to thermalize through collisions with $X,\,\bar{X}$ as $\tau_{\ell X}=\frac{3 m_{X} T_{\text{gal}}^{3 /2}}{16  e^4_D \left(2 \pi m_{\ell}\right)^{1 / 2}n_{X} \log \Lambda }$, where  $e_D$ is the charge of $\ell$, $T_{\text{gal}}\sim  \frac{1}{2}m_X v^2_{\text{gal}}$  and $\log \Lambda$ is the Coulomb logarithm,  $\Lambda\sim\frac{\sqrt{m_e T_X}}{ m_{D}}$. The resulting time is much shorter than the age of the galaxy for all of our parameter space. In any case, increasing the velocity of $\ell$s lowers the recombination rate and so Eq.~\ref{eq:rec_rate} can be seen as a conservative estimate.

The rearrangement cross section can be read directly from Eq.~\ref{eq:rea} (see also Fig. 2 of \cite{Bracci:1979iw}), with the effective ``temperature'' set to $T_{\text{\text{gal}}}=\frac{1}{2}\,m_X\,v^2_{\text{\text{gal}}}=0.1\,E_B$. Substituting this temperature, we get 
\begin{equation}
\sigma_{\text{rear}}=\frac{8.4}{\alpha^2_D\,m^2_\ell}\, .
\end{equation} 
The SDDM production rate per volume is then given by
\begin{eqnarray}
\label{eq:rear_rate}
&&\frac{d\Gamma_{\text{rear}}}{dV}~=~r_{\text{L}}\,n_{(X\ell)}\,n_{\bar{X}}\,\sigma_{\text{rear}}\,v_{\text{\text{gal}}}~=~3.5\cdot 10^{-26}\,\text{cm}^{-3}\text{s}^{-1}\nonumber\\
&&\left(\frac{n_{(X\ell)} n_{\bar{X}}}{\braket{n^{\text{total}}_\ell n_{\bar{X}}}}\right){\left(\frac{r_{\text{L}}}{0.2}\right)}\,{\left(\frac{0.1}{\alpha_D}\right)}^5\,{\left(\frac{r_{X\ell}}{m_p/m_e}\right)}^{\frac{3}{2}}\,{\left(\frac{v_{\text{\text{gal}}}}{10^{-3}\,c}\right)}\, ,\nonumber\\
\end{eqnarray}
where $r_{\text{L}}$ is the $\mathcal{O}(1)$ fraction of high-L states out of the generated $(X\bar{X})$.

\subsection{Cosmological Stability of $\left(X\bar{X}\right)$ states}
\label{sec:stability}
There are four processes that might lead to the elimination of $\left(X\bar{X}\right)$ states, namely:
\begin{itemize}
\item Direct self-annihilation of high-L states 
\item De-excitation followed by self-annihilation
\item Self-destruction through collisions with free $X$ or $\bar{X}$ particles
\item Inverse rearrangement $(X\bar{X})\,+\,\ell\rightarrow(X\ell)\,+\,\bar{X}$. 
\end{itemize}
We now show that all of these processes are slow enough such that the $(X\bar{X})$ has a phenomenologically significant abundance in the galaxy.

\subsubsection{Direct self-annihilation}

First, we consider direct self-annihilation of the high-L $(X\bar{X})$ states. This is highly suppressed due to the large centrifugal barrier preventing the wavefunctions of the $X$ and $\bar{X}$ from overlapping. Indeed, the self-annihilation rate for the $(n,L)$ state is proportional to $|\partial^L \Psi|_{r=0}|^2$, and given by \cite{An:2016gad}
\begin{equation}
\Gamma_{(n, L)} \sim\left(\frac{\alpha_{D}}{n}\right)^{2 L+3} f_\alpha\, m_{X}\, .
\end{equation}
The factor $f_\alpha$ depends on whether the bound state is $\mathcal{C}_{\text{SM}}$-even (para), or $\mathcal{C}_{\text{SM}}$-odd (ortho). In the para case, the bound state decays to $A_V\,A_V,\,A_D\,A_V$ or $A_D\,A_D$, and so $f_\alpha=\alpha^2_V+2\alpha_V\alpha_D+\alpha^2_D$. In the ortho case, it decays to $\ell^+\ell^-$, and so  $f_\alpha=\alpha^2_D/3$. The $\alpha^{2L+3}_D$ dependence comes from the wavefunction suppression in the bound state. This gives a lifetime hierarchically larger than the age of the universe. Note that the same conclusion holds when considering QED corrections to the $(X\bar{X})$, with the replacement \cite{Jaffe:1989jz} $L\rightarrow J_{\text{QED}}=L+S+J_{\text{photons}}$, so that the $J_{\text{QED}}=30$ is cosmologically stable.

\subsubsection{De-excitation}
The $(X\bar{X})$ in the universe can be eliminated by a process of spontaneous de-excitation followed by self-annihilation. This process is suppressed by the need to undergo forbidden transitions. Crucially, the rearrangement process produces highly excited $\left(X\bar{X}\right)$ states. This is because the typical impact parameter of $\bar{X}$ on $(X\ell)$ is of order $\frac{1}{\alpha_D m_\ell}\gg r^{XX}_{\text{Bohr}}$.
Consequently, $(X\bar{X})$ bound states are produced with $L\,\in\,[0,30]$ as can be see from Fig.~3 of \cite{Bracci:1979iw} at $T=T_{\text{\text{gal}}}=0.18\,E_B$. The majority of these states are produced with a high angular momentum, as the number of states scales linearly with $L$. States produced with $n>L+1$ de-excite promptly to $\left(L+1,\,L\right)$ through a cascade of allowed $\Delta L=1$ transitions with large jumps in $n$.  In contrast, states at $\left(L+1,\,L\right)$ are stable on cosmological scales, as we check below. 
 Choosing the dark photon mass to satisfy
\begin{eqnarray}
&&m_D>\Delta E^{\left(n,\,L\right)}_{\Delta n=\Delta L}~=\nonumber\\
&& \left(\frac{1}{(L+1-\Delta L)^2}-\frac{1}{(L+1)^2}\right)\,\frac{1}{4}\alpha_D^2 \,m_X\, ,
\end{eqnarray}
with $\Delta L\sim5$, it is evident  that the $\left( L+1,\,L\right)$ bound states can only de-excite through the highly forbidden $\Delta L \geq 6$ transition. In the notation of Eq.~\ref{eq:dphrange}, we have
\begin{equation}
\label{eq:stability_frac}
f_{{X\bar{X}}}=\left(\frac{1}{(L+1-\Delta L)^2}-\frac{1}{(L+1)^2}\right)\,\frac{m_X}{2m_\ell}\, ,
\end{equation}
where $f_{{X\bar{X}}}<1$ is required in order to have a viable mass range for $m_D$. We can see this is satisfied for states with $L>28$, which make out around 20\% of all the produced SDDM, i.e. $r_{\text{L}}=0.2$. The lifetime of the forbidden transition is given by
\begin{equation}
\tau_{\rm forbidden}~\sim~10^{-19}\,\text{s}\,{\left(\frac{\alpha_0}{\alpha}\right)}^{2-\Delta L}~{\left[(2\Delta L+1)!!\right]}^2~{\left(\frac{20 L}{\Delta L}\right)}^{2\Delta L}\, .
\end{equation}
This approximation is obtained by expanding $j_{\Delta L}( q r )\sim\frac{(q r)^{\Delta L}}{(2\Delta L+1)!!}$ around $q r\sim (\Delta E/2)\,a_0\,n^2$ where the overlap integral has maximal support. 
The $\Delta L \geq 6$ bottle neck protects the $\left(X\bar{X}\right)$ from the forbidden transition, with a lifetime of $\tau_{\rm forbidden}  \sim 10^{21}\,\text{s}$.

Another possible channel for de-excitation is through the emission of an off-shell $A_V$ which goes to two neutrinos by loop-level mixing with the SM Z-boson. The lifetime for this highly suppressed process was estimated in \cite{Grossman:2017qzw} to be larger than $10^{41}\,\text{s}$.

\subsubsection{Self-Destruction}
Here we consider the typical time for an $\left(X\bar{X}\right)$ to collide with a free $X$ or $\bar{X}$, lose angular momentum, and become unstable. In particular, if the $X\bar{X}$ goes to a sufficiently small $L_{\text{small}}\sim 10$, it can de-excite promptly through a cascade of spontaneous emissions, first to $(L_{\text{small}}+1,L_{\text{small}})$, and then through $\Delta L=1$ transitions, which are larger than the photon mass below $L\sim 10$. When the $\left(X\bar{X}\right)$ reaches a low enough angular momentum, it self-annihilates promptly. 
The lifetime of $\left(X\bar{X}\right)$ in the free $X$ or $\bar{X}$ plasma is given by
\begin{equation}
\tau^{X\bar{X}}_{\text{scat}} = \frac{1}{n_X\,\sigma^{X\bar{X}}_{\text{scat}} \, v_{\text{\text{gal}}}}\, ,
\end{equation}
where $\sigma^{X\bar{X}}_{\text{scat}}$ is cross section for $\left(X\bar{X}\right)$ to scatter into a state with $L<L_{\text{small}}$, i.e.
\begin{equation}
\frac{d\sigma^{X\bar{X}}_{\text{scat}} }{d|q|^2}~=~\frac{4\pi\alpha_D^2}{|q|^4v_{\text{\text{gal}}}^2} F^2_{X\bar{X}}(q)\,. \label{eq:SDDMkill}
\end{equation}
In the latter equation, $\Delta E$ is the difference in the binding energy of the initial and final state.  
The form factor $F_{X\bar{X}}$ can be calculated using:
\begin{equation}
F_{X\bar{X}}(q)~=~\int d^3 x \,\Psi^*_i(x) \Psi_f(x) \left[e^{i q x/2} -e^{-i q x/2}\right]\,.  \label{eq:formfactor}
\end{equation}
We calculate this cross section for $r_{X\ell}=m_p/m_e$ and find that $\sigma^{X\bar{X}}_{\text{scat}}=\hat{\sigma}_{\text{scat}}\,\left(\frac{\text{GeV}}{m_X}\right)^2$ with:
\begin{equation}
\hat{\sigma}_{\text{scat}}~\simeq~2.2\times 10^{-20}\,\text{ cm}^2\,,
\end{equation}
for $\alpha_D=0.1$.

The resulting lifetime is then $\Gamma^{X\bar{X}}_{\text{scat}}=1/\tau^{X\bar{X}}_{\text{scat}} $ where
\begin{eqnarray}
\label{eq:scat_lifetime}
&&\tau^{X\bar{X}}_{\text{scat}} = 3\cdot\,10^{18}~\text{s}\nonumber\\
&&{\left(\frac{\hat{\sigma}^{\alpha_D=0.1}_{\text{scat}}}{\hat{\sigma}_{\text{scat}}}\right)}{\left(\frac{\alpha_D}{0.1}\right)}^2\,{\left(\frac{m_X}{\text{GeV}}\right)}\,{\left(\frac{\sqrt{\braket{n^2_X}}}{n_X}\right)}\,{\left(\frac{10^{-3}\,c}{v_{\text{\text{gal}}}}\right)}\,,\nonumber\\
\end{eqnarray}
which is larger than the age of the universe for most of our parameter space.

\subsubsection{Inverse Rearrangement}
We assume that the cross section for inverse rearrangement $\sigma_{\text{inv-rear}} \sim \sigma_{\text{rear}}$. This is close to the unitarity bound and is in line with the full quantum result for the muon system following \cite{InvSak}, where the inverse rearrangement cross section was actually an ${\cal O}(1)$ smaller. The lifetime for inverse rearrangement is then given by
\begin{eqnarray}
\label{eq:inv_lifetime}
&&\tau^{X\bar{X}}_{\text{inv-rear}} = 1.4\cdot\,10^{16}~\text{s}\nonumber\\
&&\left(\frac{\sqrt{\braket{\left(n^{\text{total}}_\ell\right)^2}}}{n_\ell}\right){\left(\frac{\alpha_D}{0.1}\right)}^3\,{\left(\frac{m_X}{\text{GeV}}\right)}\,{\left(\frac{m_p/m_e}{r_{X\ell}}\right)^{2}}\,{\left(\frac{10^{-3}\,c}{v_{\text{\text{gal}}}}\right)}\, .\nonumber\\
\end{eqnarray}
  
This is comparable to the age of the universe for our nominal parameters, even if slightly below it. 


\subsection{The Galactic Number Density of SDDM}

To find the number density of SDDM today we solve the coupled Boltzmann equations for $n_{(X\bar{X})}$ and $n_{(X\ell)}$, taking into account that $n^{\text{free}}_\ell+n_{(X\ell)}=n^{\text{total}}_\ell$. These equations are:
\begin{eqnarray}
\frac{d n_{(X\bar{X})}}{dt}~&=&~~~ ~~~~~\frac{d\Gamma_{\text{rear}} }{dV} - \frac{n_{(X\bar{X})}}{\tau^{X\bar{X}}_{\text{inv-rear}}} - \frac{n_{(X\bar{X})}}{\tau^{X\bar{X}}_{\text{scat}} }\nonumber\\
 \frac{d n_{(X\ell)}}{dt}~&=& ~  -\frac{1}{r_\text{L}}\frac{d\Gamma_{\text{rear}} }{dV} + \frac{n_{(X\bar{X})}}{\tau^{X\bar{X}}_{\text{inv-rear}}} + \frac{d\Gamma_{\text{rec}} }{dV}\, .
\end{eqnarray}
  
We solve these equations numerically, averaged over the galactic profile, to calculate the average SDDM fraction in our galaxy today as a function of $m_X$, under the assumption that the local density of SDDM can be inferred directly from the average fraction and the incoming DM flux. This holds when the SDDM distribution follows the cuspy  NFW or the cored Burkert distribution \cite{Nesti:2013uwa}, and so should hold reasonably well for a variety of different SDDM distributions. Nevertheless, the above assumption should be verified in a more robust analysis of galactic dynamics in our setting.  

The result of this calculation is used in Fig.~\ref{FIG:SDDM_detection}, where $\Omega_{\text{SDDM}}\sim10^{-11}-10^{-7}$ (see top horizontal axis). As we will see in the next Section, despite this incredibly small fraction, the detection of these bound states is possible even with existing data. 

 \section{Detection at neutrino detectors }
\label{sec:detection}
\begin{figure*}[t]
\begin{center}
\includegraphics[width=8cm]{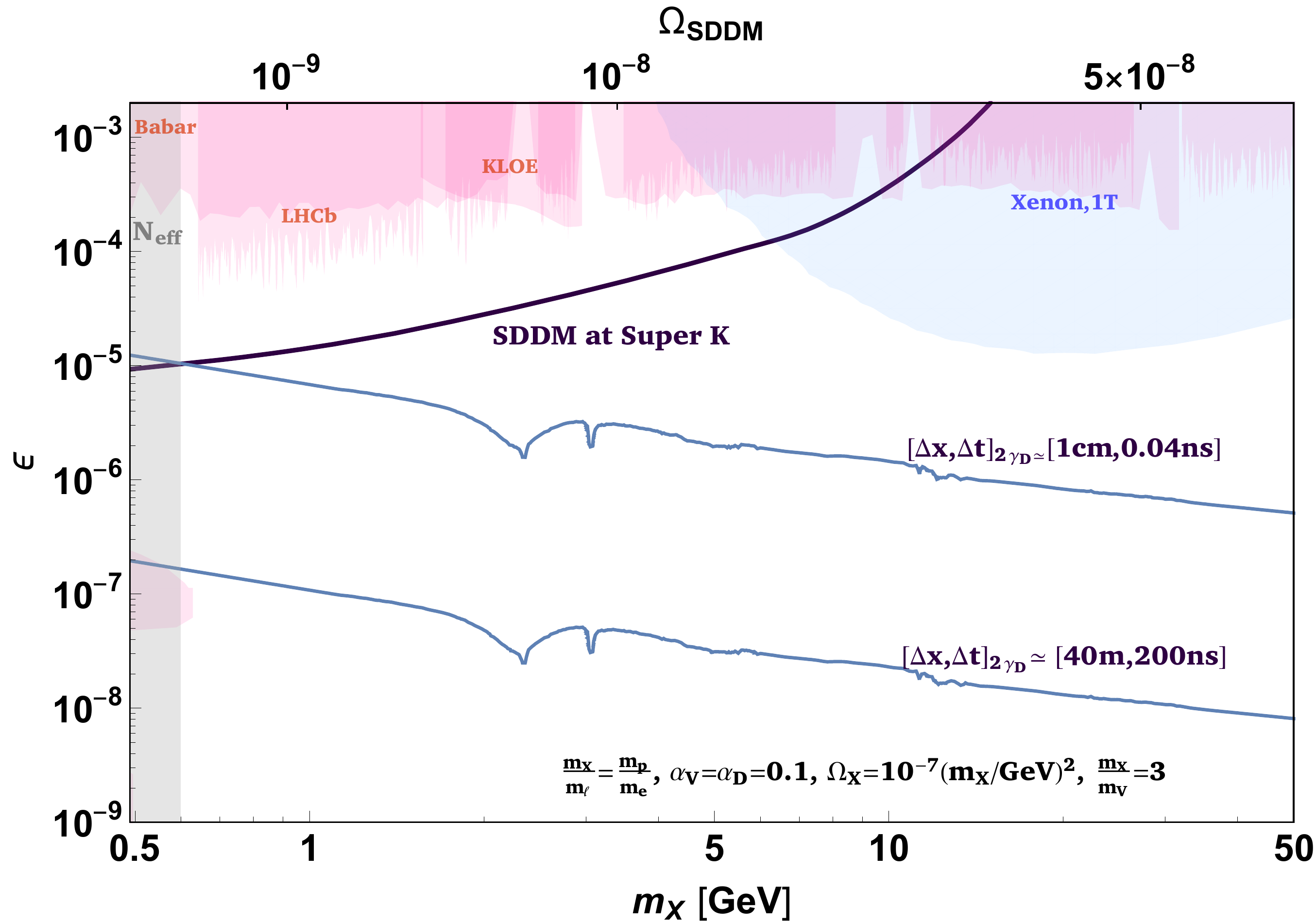}
\includegraphics[width=8cm]{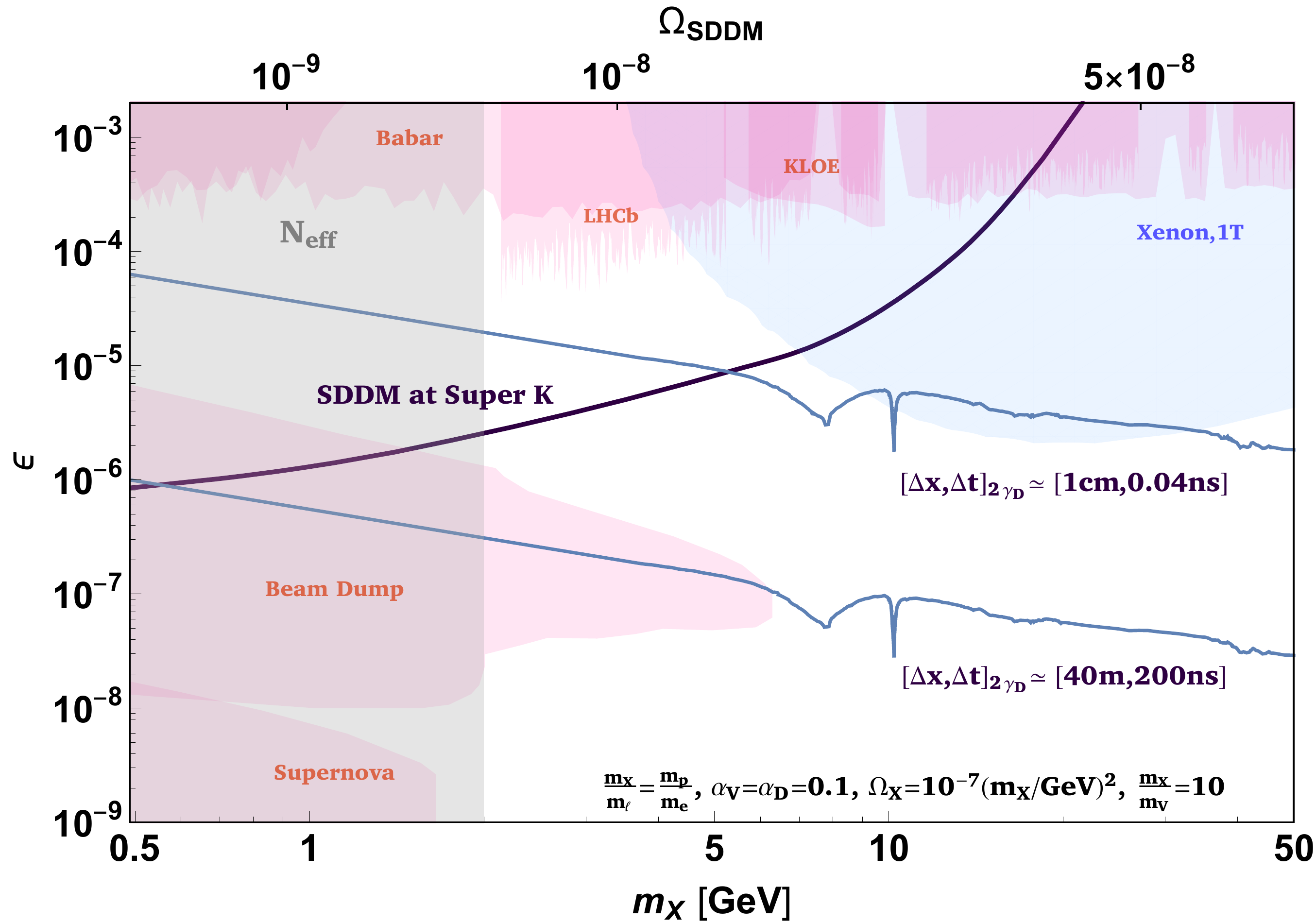}
\caption{\label{FIG:SDDM_detection} 
The 100 events/year reach for $(X\bar{X})$ SDDM (dark purple) in Super-K in the $m_X$ - $\epsilon$ plane for $m_V \,=  \,m_X/3$ (left) and for $m_V \,=\,m_X/10$ (right). The blue shaded region is the bound on free $X,\bar{X}$ from XENON1T  \cite{Aprile:2019xxb}, while the pink shaded region represents the limits on the dark photon $V$ and the darkly charged $X$ via their production in experiments - such as LHCb \cite{Aaij:2017rft,Aaij:2019bvg}, KLOE \cite{Anastasi:2018azp}, BABAR  \cite{Lees:2014xha}, Na48 \cite{Goudzovski:2014rwa}, and several beam dump experiments \cite{Bl_mlein_2014}; and also in Supernova 1987a  \cite{Chang:2018rso}. In the grey area the decoupling between the two sectors occurs after the QCD phase transition, which is disfavored by $N_{eff}$ constraints. We see that a potential dedicated analysis at Super-K could probe previously unexplored regions of parameter space. The smoking gun signal in this analysis is two simultaneous energy depositions from the self-destruction of SDDM to back-to-back dark photons, which in turn decay to SM jets or lepton pairs. The blue lines correspond to a 1cm and 40m (the size of Super-K) mean distance between the decay point of the two dark photons. The 40m line also corresponds to a mean time difference of $200\,\text{ns}$ at Super-K, indicating that the two energy depositions will count as a single event.  The blue and pink bounds are reproduced from \cite{Curtin:2014cca} and \cite{Ilten:2018crw}.\label{FIG:moneyplot}
}
\end{center}
\end{figure*}
Upon its collision with the earth through the dark photon portal, SDDM loses angular momentum, and subsequently self-annihilates into two or more dark photons. The cross section for collisional de-excitation with SM nuclei is calculated similarly to the $(X\bar{X})$ de-excitation from collisions with free $X$ (Eq.~\ref{eq:SDDMkill}). In the nucleus case, the cross section is given by:
\begin{equation}
\frac{d\sigma_{XX}}{d|q|^2} = \frac{4\pi\alpha_V \alpha }{\left(|q|^2 +m^2_V\right)^2v_{\text{\text{gal}}}^2} F_{X\bar{X}}(q) \, F_{\text{nuc}}(q)
\end{equation}
where $F_{\text{nuc}}(q)$ is the Woods-Saxon form factor \cite{Woods:1954zz}, and $F_{X\bar{X}}$ is given in Eq.~\ref{eq:formfactor}. The resulting unstable $(X\bar{X})$ state decays promptly to a number of dark photons $A_D$ or $A_V$. For direct detection purposes, we are particularly interested in the process
\begin{equation}\label{eq:rate}
(X\bar{X})\rightarrow (A_V\rightarrow \text{SM}) (A_V\rightarrow \text{SM})
\end{equation}
 with a branching ratio of $BR(X\bar{X})\rightarrow A_VA_V\sim1/4\cdot1/4$, since this decay only happens for $1/4$ of the para states, which occur $1/4$ of the time (see section~\ref{sec:stability}). This results in two energy depositions with $\mathcal{O}(m_X)$ energy, with a mean time difference $\Delta t=10\text{ps}-200\text{ns}$, depending on the dark photon lifetime and boost  (see Figure~\ref{FIG:moneyplot}). Since the original $(X\bar{X})$ is slow moving, the dark photons emerge back-to-back, and the SM energy depositions are separated by a distance of $\beta_V\Delta t\sim1\,\text{mm}-40\,\text{m}$. The SM energy depositions can be either a dilepton pair or jets. Notably, the initial collision serves merely as a trigger for SDDM self-destruction, and the SM products have $\mathcal{O}(m_X)$ energies. Consequently, the only important factor in SDDM detection is the fiducial mass of the detector and the exposure time. This makes the Super-Kamiokande detector ideal for SDDM searches, with a fiducial volume of $50\,\text{kton}$ of ultra-pure water. 

In the vast majority of our parameter space, the time difference between the decay of the two dark photons is smaller than the $200\text{ns}$ sliding window used in Super-K. The two $A_V$ decays will be counted as one single event that is very different from all other decay events in Super-K, with two primary vertices instead of one, separated by $\sim1\,\text{mm}-40\,\text{m}$. Since both vertices have similar scintillation energies associated with them, this should result in low maximum-likelihood in the standard reconstruction of the event as a single vertex event. A dedicated analysis will then require a modification of Super-K's event reconstruction algorithm \cite{Missert_2017,Jiang:2019xwn} to allow for two primary vertices. Additionally, if both dark photons decay leptonically, there will be exactly two Cherenkov rings of the same kind associated with each vertex, of similar brightness, and so it should be possible to resolve the two primary vertices in the event. These unusual characteristics could be used to conduct a very low background search for SDDM at Super-K.

In Fig.~\ref{FIG:moneyplot} we present the discovery reach for SDDM in Super-K, assuming a \textit{signal} rate of 100 events/year for mass ratios of $m_X/m_V=3$ (left panel) and $m_X/m_V=10$ (right panel). 
Additional bounds on our model are:
\begin{enumerate}
\item Standard direct detection of DM~-~nuclear and electron recoil at XENON1T \cite{Aprile:2019xxb}. This bound is shown in Fig.~\ref{FIG:moneyplot} shaded in blue. 
\item Bounds on production of $X$ and $A_V$~-~these bounds are shaded in light red in Fig.~\ref{FIG:moneyplot}. The bounds include SN1987A cooling \cite{Chang:2018rso},  LHCb \cite{Aaij:2017rft,Aaij:2019bvg}, KLOE \cite{Anastasi:2018azp}, BABAR  \cite{Lees:2014xha}, Na48 \cite{Goudzovski:2014rwa}, and several beam dump experiments \cite{Bl_mlein_2014}. The relevant bounds were reproduced from \cite{Curtin:2014cca} and \cite{Ilten:2018crw}.
\end{enumerate}

We see that the discovery prospects in the existing data are not currently constrained by any other experiments in a large fraction of the parameter space. This is particularly striking in light of the low values of $\Omega_{\text{SDDM}}$ depicted in the plot. The reason for this is of course the large volume of the Super-K detector.

\section{Outlook}

In this paper we introduced a model of self-destructing DM with a viable cosmological history. The dark sector has a component of mostly symmetric free heavy fermions $X,\,\bar{X}$ and a small asymmetric population of light $\ell$ fermions. We assume conservatively that the early universe dynamics leaves all of the $\ell$ free.
In the higher density environment of the galaxy, recombination into $(X\ell)$ atoms is constantly occurring, followed by an atomic rearrangement reaction $(X\ell)+\bar{X}\rightarrow (X\bar{X})+\ell$ analogous to muon capture in the SM. The latter process was studied by Fermi and Teller in the `40s, and found to be geometric and adiabatic. Furthermore, the atomic rearrangement process in our model naturally generates high angular momentum $(X\bar{X})$, which is our SDDM candidate. 

The high-L $(X\bar{X})$ have a lifetime comparable to the age of the universe - they are protected from self-annihilation by virtue of their large angular momentum, and from spontaneous emission due to the mass of the dark photon, which implies highly forbidden radiative transitions. 

When incident upon the detector, the $(X\bar{X})$ can undergo collisional de-excitation with the nucleus through the dark photon portal, transitioning into an unstable low-L $(X\bar{X})$ state. The resulting low-L state is no longer protected by angular momentum, and so it self-annihilates promptly into two or more dark photons, which decay in the detector, leading to striking signals in large volume, high threshold neutrino detectors. For example, 
this could lead to smoking-gun signatures in the Super-Kamiokande detector, involving simultaneous $\ell^\pm$ pairs, a macroscopic distance apart. We explore the detection prospects for such a signal in Super-K, and find that it can be discovered in a dedicated analysis of the existing data. 

\mysections{Acknowledgments}
 We thank Asher Berlin, Itay Bloch-Mimouni, Jeff Dror, Mitrajyoti Ghosh, Yann Gouttenoire, Yuval Grossman, Ed Kearns, Simon Knapen,  Eric Kuflik,  Marcus Luty, Nadav Joseph Outmezguine, Yael Shadmi, Yotam Soreq, Chen Sun,  and Tomer Volansky for useful discussions. We would especially like to thank Jeff Dror, Eric Kuflik and Tomer Volansky for invaluable comments on the draft. Many thanks to Asher Berlin and Simon Knapen, who helped us avoid an unnecessary mistake concerning the BBN bound on our model. MG thanks the hospitality of KITP at UC Santa Barbara and MIAPP where part of this project was completed. MG is supported in part by the Israel Science Foundation (Grant No. 1302/19). OT is supported in part by the DOE under contract DE-AC02-05CH11231.

\bibliography{SDDM}{}

\end{document}